# High anisotropy of fully hydrogenated borophene

Zhiqiang Wang,[1] Tie-Yu Lü,[1] Hui-Qiong Wang,[1,2] Yuan-Ping Feng[3], Jin-Cheng Zheng,[1,2,4*]

[1] Department of Physics, and Collaborative Innovation Center for Optoelectronic Semiconductors and Efficient Devices, Xiamen University, Xiamen 361005, China

[2] Xiamen University Malaysia, 439000 Sepang, Selangor, Malaysia

[3] Department of Physics, National University of Singapore, Singapore 117542, Singapore

[4] Fujian Provincial Key Laboratory of Theoretical and Computational Chemistry, Xiamen University, Xiamen 361005, China

* Author to whom correspondence should be addressed. Electronic mail: jczheng@xmu.edu.cn

**Abstract**

We have studied the mechanical properties and phonon dispersions of fully hydrogenated borophene (borophane) under strains by first principles calculations. Uniaxial tensile strains along the *a*- and *b*-direction, respectively, and biaxial tensile strain have been considered. Our results show that the mechanical properties and phonon stability of borophane are both highly anisotropic. The ultimate tensile strain along the *a*-direction is only 0.12, but it can be as large as 0.30 along the *b*-direction. Compared to borophene and other 2D materials (graphene, graphane, silicene, silicane, h-BN, phosphorene and $MoS_2$), borophane presents the most remarkable anisotropy in in-plane ultimate strain, which is very important for strain engineering. Furthermore, the phonon dispersions under the three applied strains indicate that borophane can withstand up to 5% and 15% uniaxial tensile strain along the *a*- and *b*-direction, respectively, and 9% biaxial tensile strain, indicating that mechanical failure in borophane is likely to originate from phonon instability.

## Introduction

Two-dimensional (2D) boron sheet (borophene) has attracted much attention[1-6] since it was synthesized on a silver substrate under ultrahigh-vacuum[7]. While experimental progress is slow due to difficulty in synthesizing borophene under the strict experimental conditions, many first principles calculations have been performed to study the mechanical properties, electronic structure, phonon dispersion, lattice thermal conductivity, superconducting behavior and optical properties of borophene, as well as borophene nanoribbons and borophene nanotubes[8-19]. It has been shown that the mechanical properties of borophene are highly anisotropic[8]. The ultimate strains along the *a*-direction and the *b*-direction (zigzag-direction) are 0.08 and 0.15, respectively. The buckling height of borophene is 0.91 Å. When a biaxial tensile strain is applied to borophene, the buckling height decreases with the strain and becomes zero when the strain reaches 0.13. The superconducting transition temperature $T_c$ of borophene is about 19 K, which can be increased to 27.4 K by applying a tensile strain, or 34.8 K by hole doping[9]. Due to the strong phonon-phonon scattering, the lattice thermal conductivity of borophene is unexpectedly low[20]. Recently, Feng *et al.* synthesized two boron sheets: $β_{12}$ and $χ_3$, both of which belong to the triangular lattice but differ in arrangements of periodic holes[21]. Both $β_{12}$ and $χ_3$ boron sheets are found planar without vertical buckling. In addition, the potential of borophene as an anode material for lithium, sodium and magnesium ion batteries has been investigated. The theoretical specific capacities are 1860, 1218 and 1960 mAh/g for lithium, sodium and magnesium ion battery, respectively. It is very interesting that the energy barriers of lithium, sodium and magnesium diffusion along the *a*-direction are only 2.6, 1.9 and 28 meV, indicating that lithium, sodium and magnesium can fast diffuse along the *a*-direction at room temperature[22-24].

Hydrogenation is an important approach to modify the physical and chemical properties of 2D materials[25, 26]. Different hydrogenation patterns on the same 2D material can lead to different physical and chemical properties of the material[27, 28]. For instance, graphene is a zero-gap semiconductor. The band gap of graphene can be tuned into a wide range of values by controlling hydrogen coverage. The band gap of graphene which is fully hydrogenated on one side is about 1.2 eV[29], but that fully hydrogenated on both sides (graphane) becomes 3.5 eV[30]. Therefore, hydrogen coverage strongly influences the optical properties of graphene[31]. Furthermore, at very low concentration of hydrogen, some interesting phenomena such as hydrogen-induced ferromagnetism can be observed[32]. For BN sheet, the band gap can be narrowed significantly by hydrogenation[33]. Hydrogenation also changes the mechanical properties of graphene and BN sheet significantly[34, 35]. For graphene, the Young's modulus changes from 354 N/m in pristine graphene to 248 N/m in fully hydrogenated graphene[36].

It has been reported that full hydrogenation can further stabilize borophene. The fully hydrogenated borophene, or Borophane, has a direction-dependent Dirac cone, and the Dirac fermions possess an ultrahigh Fermi velocity (3.0×10⁶ m/s). This, combined with the excellent mechanical performance of borophane, makes borophane attractive for applications in nanoelectronics devices[37]. The mechanical property of a material is an important parameter for the application of the material[38]. For example, the knowledge of mechanical properties servers as a guidance in strain engineering, which is an important way to tune the physical and chemistry properties of materials[39-46]. To date, the strain-dependent mechanical properties of borophane have not been reported, and would be the focus of the present study.

In this work, the mechanical properties and phonon dispersions of borophane subjected to various strains are studied. Specifically, three different types of strain, (i) a uniaxial tensile strain along the $a$-direction ($\varepsilon_a$), (ii) a uniaxial tensile strain along the $b$-direction ($\varepsilon_b$), and (iii) a biaxial tensile strain in the borophane plane ($\varepsilon_{ab}$) are considered. Our results show that the mechanical properties and phonon stability of borophane are both highly anisotropic. Furthermore, the elastic properties of borophane along arbitrary directions are discussed and compared with those of other 2D boron materials, borophene, α sheet, $\beta_{12}$, $\chi_3$, pmmm and pmmn 2D.

## Computational details

All calculations are performed using the Quantum-Espresso package[47]. Ultrasoft pseudopotentials[48] are used for all atoms and the exchange-correlation approximation is described by the Perdew-Burke-Ernzerh[49] functional. The kinetic-energy cutoff of plane-waves is set to be 50 Ry. The mesh for $k$-point sampling is 13×11×1 for the unit cell which contains two B atoms and two H atoms (see figure 1). The atomic positions and lattice constants are fully relaxed. In phonon dispersion calculation, a finer $k$-point sampling mesh of 35×25×1 is used. The forces on all atoms are less that $10^{-5}$ eV/Å. A large vacuum region (22 Å) is included in the $z$-direction to eliminate the interlayer interactions.

The four nonzero elastic constants of a 2D material are $c_{11}$, $c_{22}$, $c_{12}$ and $c_{66}$. The layer modulus is[50]

$$\gamma = \frac{1}{4}(c_{11} + c_{22} + 2c_{12}), \quad (1)$$

the Young's modulus along the $a$- and $b$-direction can be expressed as[50]

$$Y_a = \frac{c_{11}c_{22} - c_{12}^2}{c_{22}} \text{ and } Y_b = \frac{c_{11}c_{22} - c_{12}^2}{c_{11}}, \quad (2)$$

respectively, the Poisson's ratio along the $a$- and $b$-direction can be given as

$$v_a = \frac{c_{12}}{c_{22}} \text{ and } v_b = \frac{c_{12}}{c_{11}}, \quad (3)$$

respectively, while the shear modulus can be written as

$$G = c_{66}. \quad (4)$$

## Results and discussion

### 1. Unstrained borophane

The crystal structure of borophane is displayed in figure 1. The unit cell is marked by the black dashed rectangle which contains two B and two H atoms. Each B atom is hydrogenated with an H atom at the top site. The optimized lattice constants are 1.941 Å and 2.815 Å along the $a$- and $b$-direction, respectively, which are in good agreement with previous theoretical results[37]. Borophane has a buckled configuration and the buckling height is 0.81 Å, which is smaller than that of borophene (0.91 Å)[8]. The two B atoms, labeled B1 and B2 in figure 1, in the unit cell are inequivalent. For convenience of discussion, we define the bond between two B1 atoms, or equivalently two B2 atoms, in adjacent unit cells as bond1, and that between the B1 and B2 atoms as bond2. For strain-free borophane, the lengths of bond1 ($r_1$) and bond2 ($r_2$) are 1.941 and 1.890 Å, respectively. The inequivalence of bond1 and bond2 could be the cause of high anisotropy of borophane. The calculated electronic band structure, electronic density of states, phonon dispersion and phonon density of states of borophane are displayed in figure 2. The band structure of borophane shows clearly a Dirac cone along the Γ-X direction, which agrees well with previous theoretical results[37]. No imaginary frequencies were found along the high-symmetry directions of the Brillouin zone, indicating that borophane is stable. In contrast, for borophene, there exists a small imaginary frequency along the Γ-X direction, which indicates that borophene is unstable against long-wavelength transversal waves[37]. Thus, under normal temperature and pressure, borophene can be stabilized by hydrogenation.

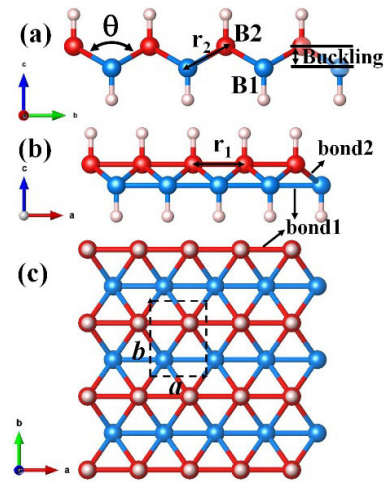

Fig. 1. Crystal structure of borophane. (a) side view with *b-c* plane shown, (b) side view with *a-c* plane shown, and (c) top view of atomic structure of borophane. The unit cell, marked by the dashed rectangle, contains two B atoms and two H atoms. The dihedral angle (θ), B-B bond and bond length $r_1$, and $r_2$, as well as the buckling distance are shown in (a) and (b). The big red and blue balls represent B atoms, the small light pink balls represent H atoms, respectively.

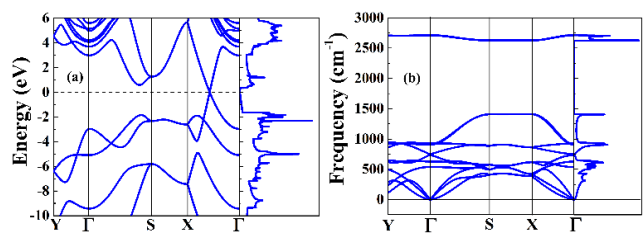

Fig 2. (a) Electronic band structure and density of states of borophane. (b) Phonon dispersion and phonon density states of borophane. The Fermi energy is set at 0 eV.

## 2. Mechanical properties

The stress-strain curves of borophane under various strains considered are displayed in figure 3. First of all, the mechanical properties of borophane are highly anisotropic. Under a uniaxial strain in the *a*-direction, the stress increases with increasing strain until $\varepsilon_a$ = 0.12, beyond which the stress decreases sharply. This is an indication that borophane is unstable once the uniaxial strain in the *a*-direction reaches 0.14. The stress-strain curve under the uniaxial strain in the *b*-direction is homologous with that under the biaxial strain. The ultimate tensile strains are 0.12, 0.30 and 0.25 for uniaxial strain along the *a*- and *b*-direction, and biaxial strain, respectively. The corresponding ideal strengths are 12.26, 18.48 and 21.06 N/m. Borophane possess superior mechanical flexibility along the *b*-direction, which will be further discussed below. The high ultimate strains make borophane potentially useful in high-strain engineering applications.

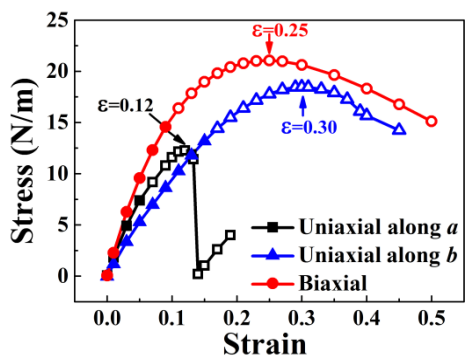

Fig. 3. Calculated stress-strain curve of borophane under uniaxial strain along the *a*- and *b*-direction, respectively and biaxial strain. Borophane can sustain stresses up to 12.26 N/m, 18.48 N/m and 21.06 N/m in the *a*, *b*, and biaxial directions, respectively. The corresponding critical strains are 0.12 (*a*-direction), 0.30 (*b*-direction), and 0.25 (biaxial direction).

In order to understand the highly anisotropic mechanical property, we have calculated the buckling heights, dihedral angles and B-B bond lengths under the three types of applied strains and show the results in figure 4. Buckling height is a critical parameter for buckled 2D material. As seen in figure 4 (a), under the uniaxial strain in the *a*-direction, the buckling height increases gradually with strain until $\varepsilon_a$ = 0.12 beyond which there is a sudden jump, which coincides with the inflexion in the tress-strain curve. The variation of buckling height under the uniaxial strain along the *b*-direction is homologous with that under the biaxial strain for $\varepsilon_b$<0.35. In both cases, the buckling of borophane maintains well. At $\varepsilon_b$ = 0.3 and $\varepsilon_{ab}$ = 0.3, the buckling heights are 0.74 Å and 0.72 Å, respectively, which correspond to only 8.6% and 11.1% decrement compared to that of strain-free borophane. In contrast, it was reported that the buckling height of borophene reduces to zero when the biaxial tensile strain reaches $\varepsilon_{ab}$ = 0.13[8]. Similarly, the buckling height of silicene drops to zero when a uniaxial strain along the zigzag direction reaches 0.17[51]. However, the buckling height of fully hydrogenated silicene remains at 0.63 Å at the same strain[52]. These results indicate that hydrogenation is helpful for maintaining the buckling height under tensile strains because hydrogenation can stabilize $sp^3$ hybridization. The variation of the dihedral angle θ, defined in figure 1, with strain shows a reverse trend, compared with the buckling-strain curve. However, it is in good agreement with the result of the buckling height variation. For unstrained borophane, the dihedral angle θ is 120.44º. At $\varepsilon_b$ = 0.3, the dihedral angle θ increases to 134.64º. Similarly, at $\varepsilon_{ab}$ = 0.3, the dihedral angle θ increases to 134.66º. This indicates that the uniaxial tensile strain along the *b*-direction and the biaxial strain flatten the configuration of borophane.

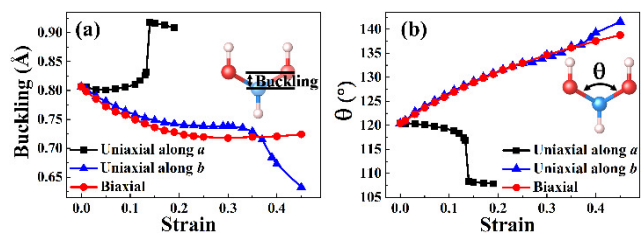

Fig. 4. Strain-dependent buckling (a) and dihedral angle θ (b) of borophane, for various types of strains considered.

To further analyze the structural changes of borophane under the three types of strains considered, we show in figure 5 the strain induced change in bond length relative to that in unstrained borophane, $\Delta r/r_0$, where $\Delta r$ is the change of B-B bond length due to applied strain, and $r_0$ is the B-B bond length in unstrained borophane. Under a uniaxial strain in the *a*-direction, the length of bond1 increases with increasing strain appreciably and linearly, while bond2 is elongated slightly until $\varepsilon_a$ = 0.13. The $\Delta r/r_0$ - $\varepsilon_a$ curve shows an inflexion point around 0.15. On the other hand, under the uniaxial strain in the *b*-direction, the length of bond1 decreases slowly while the length of bond2 increases with strain. The percentage change in the length of bond1 is 12% at $\varepsilon_a$ = 0.12, compared to 6% change in the length of bond2 at the same strain value $\varepsilon_b = 0.12$ in the *b*-direction. Due to buckling along the *b*-direction, the strain induced change in B-B bond length is much smaller in the *b*-direction compared to that in the *a*-direction. This is understandable since under a uniaxial strain in the *b*-direction, the increase in the dihedral angle θ helps to release the tensile strain, and on the other hand, bond1 is parallel to the *a*-direction and is therefore most stretched under a uniaxial tensile strain in the *a*-direction. Due to the angle (41.88°) between bond2 and the *b*-direction, a uniaxial tensile strain in the *b*-direction stretches the pucker of borophane, rather than significantly extending the B-B bond lengths.

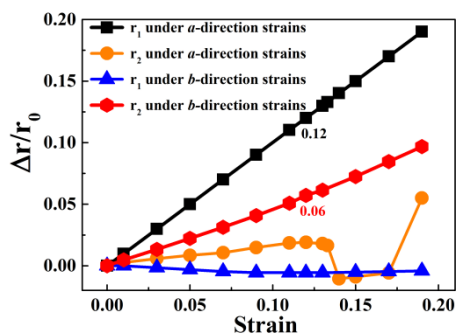

Fig. 5. The B-B bond length difference with respect to unstrained bond, $\Delta r/r_0$, as a function of uniaxial strain along the *a*- and *b*-direction strain, respectively. $\Delta r$ is the strain induced change of B-B bond length and $r_0$ is the B-B bond length in unstrained borophane.

Here, we discuss the anisotropy of borophane as well as other 2D materials in terms of ultimate strains along the armchair (*a*-direction) and the zigzag (*b*-direction) directions. The ratio ($\eta^{ZZ}/\eta^{AC}$) of the ultimate strains along the zigzag ($\eta^{ZZ}$) and the armchair ($\eta^{AC}$) direction is calculated for borophane and is shown in figure 6, together with those for graphene[53], graphane[54], borophene[8], silicene[55], silicane[56], h-BN[57], phosphorene[58] and MoS2[59]. Compared with other 2D materials, borophane has a much higher $\eta^{ZZ}/\eta^{AC}$. In particular, the ratio of the ultimate strains in borophane (2.5) is much larger compared to that in borophene (1.5). The huge difference in ultimate strains along the armchair and the zigzag directions of borophane indicates that uniaxial strain along the zigzag direction has priority over the uniaxial strains along the armchair direction in strain engineering. This can be attributed to two factors: (1) Borophane shows superior mechanical stability along the zigzag direction. (2) Due to the compensation of the dihedral angle and B-B bond angles, the strain energy in the zigzag direction is less than that along the armchair direction. As a result, it is easier to apply strain along the zigzag direction. It is very interesting that the ultimate strain under a biaxial strain (0.25) is much larger than that under a uniaxial strain in the armchair direction (0.12). This implies that, under a uniaxial strain in the armchair direction, lattice expansion along the zigzag direction can help to improve the mechanical stability of borophane. This is similar to the situation in graphene which has ultimate strains of 0.17 and 0.25 along the armchair and the zigzag direction, respectively. However, the ultimate strain of graphane under a biaxial strain can reach 0.23, which is much larger than that along the armchair direction.

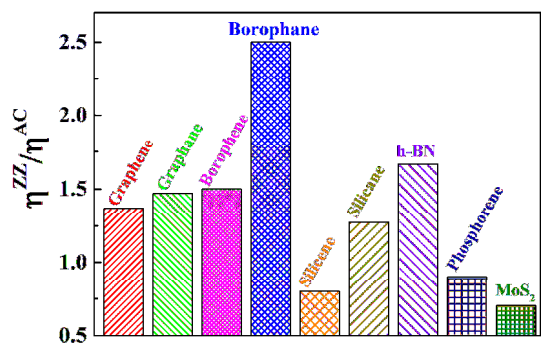

Fig. 6. The ratio ($\eta^{ZZ}/\eta^{AC}$) of ultimate strains along the zigzag ($\eta^{ZZ}$) and the armchair ($\eta^{AC}$) direction in graphene, graphane, borophene, borophane, silicene, silicane, h-BN, phosphorene and MoS2.

In addition to the ideal strength and ultimate strain, we calculated the elastic constants, layer modulus, Young's modulus, shear modulus, and Poisson's ratios of borophene and borophane. In order to see more clearly the unique properties of borophane, we also include several related materials such as α sheet, $\beta_{12}$, $\chi_3$, *pmmm* and *pmmn* 2D boron materials and graphene for comparison. Our results are in good agreement with previous theoretical results.[60, 61] The crystal structures of α sheet, $\beta_{12}$, $\chi_3$, *pmmm* and *pmmn* 2D boron materials[62] are displayed in figure 7. The α sheet, $\beta_{12}$ and $\chi_3$ boron sheets are planar without vertical buckling. The total energy of *pmmn*, α sheet, $\chi_3$ and $\beta_{12}$ are 35, 77, 110 and 123 meV/atom higher than that of *pmmm* 2D boron, which are in good agreement with previous theoretical results[62]. From the view of total-energy, with different B atom defect concentrations, $\beta_{12}$, $\chi_3$ and α sheet are more stable than borophene, respectively. It indicates that the stability of borophene can be improved by B atom defects. For silicene, the structural stabilities with typical point defects have been reported.[63] As listed in Table 1, the Young's modulus of borophene along the *a*- and *b*-direction are 378.97 and 162.49 N/m, in good agreement with previous results. When fully hydrogenated, the Young's modulus along the *a*- and *b*- direction are reduced to 172.24 and 110.59 N/m. A comparison of the B-B bond lengths in borophene and borophane reveals that the B-B bond along the *a*-direction (bond1) is elongated by 0.325 Å due to hydrogenation, leading to decreased strength of B-B bonds along the *a*-direction. As a consequence, the Young's modulus along the *a*-direction is reduced dramatically. Compared with the Young's modulus of graphene which is isotropic, the Young's modulus of both borophene and borophane are highly anisotropic. For borophene, the ratio ($Y_a/Y_b$) of Young's modulus along the *a*- and *b*- direction is 2.33 which is reduced to 1.56 in borophane due to hydrogenation. The Young's modulus of α sheet, $\beta_{12}$ and $\chi_3$ 2D boron materials along the *a*- and *b*-direction range from 179.00 to 210.56 N/m. Due to the crystal symmetry, the elastic properties of α sheet boron is isotropic. The Young's modulus and Poisson ratio of the α sheet boron is 210.56 N/m and 0.196. It is interesting to note that the Young's modulus along the *b*-direction of *pmmm* 2D boron material is considerably large (574.61 N/m), even much larger than that of graphene (338.08 N/m).

We also calculated the mechanical properties of borophene, borophane and the other five 2D boron materials along an arbitrary direction, and the results are shown in figure 8. For isotropic materials, the Young's modulus and shear modulus are independent of the direction. The polar diagrams of Young's modulus and shear modulus are perfect circles. The larger degree of deviation from a perfect circle, the stronger anisotropy the materials possess. It can be seen in figure 8 that the Young's modulus and shear modulus of α sheet, $\beta_{12}$ and $\chi_3$ 2D boron materials are not strongly dependent on the direction. But the Young's modulus of borophene, borophane, *pmmm* and *pmmn* 2D boron materials are highly anisotropic. The shear modulus of borophane is highly anisotropic, and the ratio of maximum and minimum shear modulus is 2.08, compared to 1.32 of borophene.

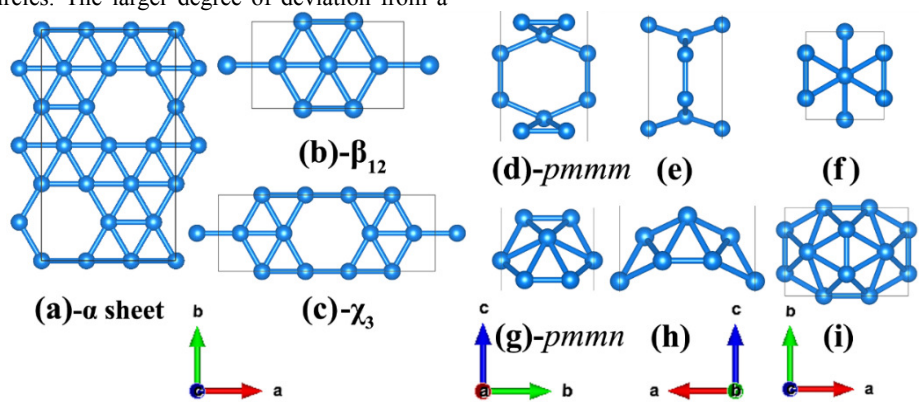

Fig. 7. Crystal structures of α sheet, $\beta_{12}$, $\chi_3$, *pmmm* and *pmmn* 2D boron materials. Top views of (a) α sheet, (b) $\beta_{12}$, (c) $\chi_3$ 2D boron materials. (d) Side view with *b-c* plane, (e) side view with *a-c* plane, and (f) top view of *pmmm* boron. (g) Side view with *b-c* plane, (h) side view with *a-c* plane, and (i) top view of *pmmn* boron.

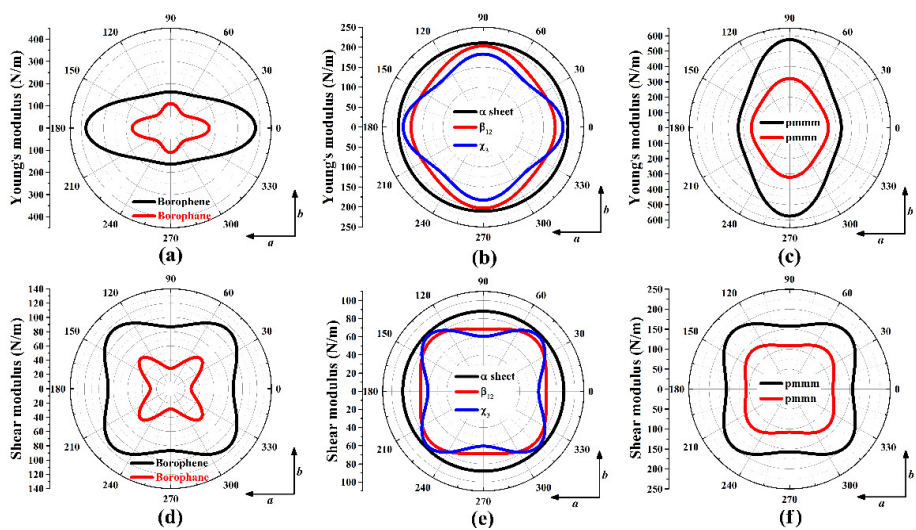

Fig. 8. Polar diagrams for Young's modulus and shear modulus of borophene, borophane, α sheet, $\beta_{12}$, $\chi_3$, *pmmm* and *pmmn* 2D boron materials. The angle is measured relative to the *a*-direction. Isotropic (anisotropic) behavior is associated to a circular (noncircular) shape.

**Table 1**. Elastic constants $c_{ij}$, layer modulus $\gamma$, shear modulus $G$, Young's modulus $Y$ in N/m, and Poisson's ratio $v$ of borophene, borophane, α sheet, $\beta_{12}$, $\chi_3$, *pmmm*, *pmmn* 2D boron materials and graphene, respectively.

|  | $c_{11}$ | $c_{22}$ | $c_{12}$ | $c_{66}=G$ | $\gamma$ | $Y_a$ | $Y_b$ | $v_a$ | $v_b$ |
|---|---|---|---|---|---|---|---|---|---|
| **Borophene** | 379.00 | 162.50 | -2.00 | 87.00 | 134.37 | 378.97 | 162.49 | -0.012 | -0.005 |
| Ref. 8 | 398.00 | 170.00 | -7.00 | 94.00 | 138.50 | 398.00 | 170.00 | -0.040 | -0.020 |
| **Borophane** | 175.77 | 112.86 | 19.97 | 28.46 | 82.14 | 172.24 | 110.59 | 0.177 | 0.114 |
| α sheet | 219.00 | 219.00 | 43.00 | 88.00 | 131.00 | 210.56 | 210.56 | 0.196 | 0.196 |
| $\beta_{12}$ | 185.50 | 210.50 | 37.00 | 68.50 | 117.50 | 179.00 | 203.12 | 0.176 | 0.199 |
| $\chi_3$ | 201.00 | 185.00 | 21.50 | 60.50 | 107.25 | 198.50 | 182.70 | 0.116 | 0.107 |
| *pmmm* | 333.50 | 576.00 | 21.50 | 157.00 | 238.13 | 332.70 | 574.61 | 0.037 | 0.064 |
| *pmmn* | 249.00 | 322.50 | 15.50 | 108.00 | 150.63 | 248.26 | 321.54 | 0.048 | 0.062 |
| Graphene | 348.75 | 348.75 | 61.00 | 143.88 | 204.87 | 338.08 | 338.08 | 0.175 | 0.175 |
| Ref. [50] | 352.70 | 352.70 | 60.90 | 145.90 | 206.80 | 342.18 | 342.18 | 0.173 | 0.173 |
| Ref. [64] | 358.10 | 358.10 | 60.40 | 148.90 | 209.25 | 347.91 | 347.91 | 0.169 | 0.169 |

## 3 Phonon dispersions of strained borophane

Phonon dispersion is important for estimating stability of crystal structure. Imaginary frequencies along any high-symmetry direction of the Brillouin zone are indications of instability of the crystal structure. The calculated phonon dispersions of borophane under the three types of applied strains are shown in figure 9. Under the uniaxial strain in the *a*-direction, two modes along the S-X direction softened and their frequencies become imaginary when the strain reaches a critical value (between 0.05 and 0.07). Under the uniaxial strain in the *b*-direction, the frequencies of long-wavelength phonons along the X-Γ direction decrease with increasing strain, and are imaginary at $\varepsilon_b$=0.17. Under the biaxial strain, imaginary frequencies along the Γ-S-K direction appear at $\varepsilon_{ab}$=0.11. The phonon dispersions of borophane under tensile strains indicate that borophane can withstand up to 5%, 15% uniaxial tensile strains along the *a*- and *b*-direction, respectively and 9% biaxial tensile strain. It is generally accepted that imaginary frequencies of phonon dispersion are signs of instability. Hence, borophane would become unstable before reaching the ultimate tensile strains. The phonon instability of borophane is highly anisotropic. The strain that borophane can withstand along the *b*-direction is 3 times of that along the *a*-direction. The phonon stability of borophane under a biaxial tensile strain is superior to that under a uniaxial strain along the *a*-direction. It implies that, the lattice expansion along the *b*-direction can significantly improve the phonon stability of borophane. Overall, borophane shows superior mechanical stability and phonon stability along the *b*-direction.

## Conclusions

A first principles study has been performed to investigate the mechanical properties and phonon stability of borophane. Our results show that the mechanical properties, and phonon stability of borophane are highly anisotropic. Compared with borophene, graphene, graphane, silicene, silicane, h-BN, phosphorene and $MoS_2$, borophane presents the most remarkable anisotropy in in-plane ultimate strain, which is very important for strain engineering. The ultimate tensile strains are 0.12, 0.30 and 0.25 for uniaxial tensile strain along the *a*- and *b*-direction and biaxial tensile strain, respectively. The phonon dispersions of borophane calculated under the various applied strains indicate that borophane can withstand up to 5%, 15% uniaxial tensile strain along the *a*- and *b*-direction, respectively, and 9% biaxial tensile strain. The failure mechanism of borophane is mainly through phonon instability. The elastic properties of borophene, borophane and related 2D boron materials along an arbitrary direction were also discussed. Our results indicate that borophane has superior mechanical flexibility and phonon stability along the b-direction.

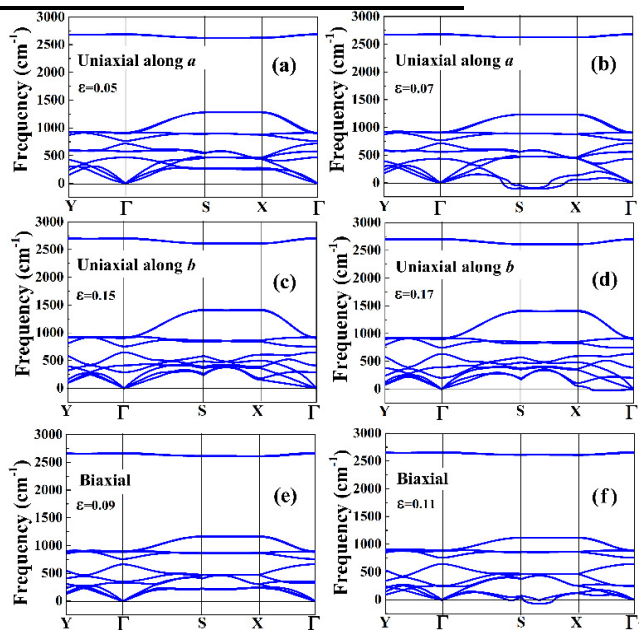

Fig. 9. The phonon dispersion of borophane under a uniaxial strain of (a) ε=0.05, (b) ε=0.07 in the *a*-direction; under a uniaxial strain of (c) ε=0.15, (d) ε=0.17 in the *b*-direction; and under a biaxial strain of (e) ε=0.09, (f) ε=0.11, respectively.

## Acknowledgements

This work is supported by the Fundamental Research Funds for Central Universities (Grant Nos. 2013121010, 20720160020), the Natural Science Foundation of Fujian Province, China (Grant No. 2015J01029), Special Program for Applied Research on Super Computation of the NSFC-Guangdong Joint Fund (the second phase), the National Natural Science Foundation of China (no. U1332105, 11335006), and the National High-tech R&D Program of China (863 Program, No. 2014AA052202). TYL acknowledges the National University of Singapore for hosting his visit during which part of the work reported here was carried out.

## Notes and references

1  E. S. Penev, A. Kutana and B. I. Yakobson, *Nano Lett.*, 2016, **16**, 2522-2526.
2  S. G. Xu, Y. J. Zhao, J. H. Liao, X. B. Yang and H. Xu, Nano Res., 2016,9(9), 2616-2622.
3  A. Lopez-Bezanilla and P. B. Littlewood, *Phys. Rev. B*, 2016, **93**, 241405(R).
4  B. Peng, H. Zhang, H. Shao, Y. Xu, R. Zhang and H. Zhu, *ArXiv e-prints 1601*, 2016, http://arxiv.org/abs/1601.00140.
5  V. Wang and W. T. Geng, *ArXiv e-prints*, 2016, **1607**, http://arxiv.org/abs/1607.00642.
6  J. Carrete, W. Li, L. Lindsay, D. A. Broido, L. J. Gallego and N. Mingo, *Mater. Res. Lett.*, 2016, DOI:

10.1080/21663831.2016.1174163, 1-8.
7  A. J. Mannix, X. F. Zhou, B. Kiraly, J. D. Wood, D. Alducin, B. D. Myers, X. Liu, B. L. Fisher, U. Santiago, J. R. Guest, M. J. Yacaman, A. Ponce, A. R. Oganov, M. C. Hersam and N. P. Guisinger, *Science*, 2015, **350**, 1513-1516.
8  H. Wang, Q. Li, Y. Gao, F. Miao, X.-F. Zhou and X. G. Wan, *New J. Phys.*, 2016, **18**, 073016.
9  R. C. Xiao, D. F. Shao, W. J. Lu, H. Y. Lv, J. Y. Li and Y. P. Sun, *ArXiv e-prints 1604*, 2016, http://arxiv.org/abs/1604.06519.
10 M. Gao, Q. Z. Li, X. W. Yan and J. Wang, *ArXiv e-prints 1602*, 2016, http://arxiv.org/abs/1602.02930.
11 Y. X. Liu, Y. J. Dong, Z. Y. Tang, X. F. Wang, L. Wang, T. J. Hou, H. P. Lin and Y. Y. Li, *J. Mater. Chem. C*, 2016, **4**, 6380-6385.
12 X. Yang, Y. Ding and J. Ni, *Phys. Rev. B*, 2008, **77**, 041402(R).
13 Z. Pang, X. Qian, R. Yang and Y. Wei, *ArXiv e-prints 1602*, 2016, http://arxiv.org/abs/1602.05370.
14 A. D. Zabolotskiy and Y. E. Lozovik, *ArXiv e-prints 1607*, 2016, http://arxiv.org/abs/1607.02530.
15 J. Yuan, L. W. Zhang and K. M. Liew, *RSC Adv.*, 2015, **5**, 74399-74407.
16 F. Meng, X. Chen and J. He, *ArXiv e-prints 1601*, 2016, http://arxiv.org/abs/1601.05338.
17 H. Liu, J. Gao and J. Zhao, *Sci. Rep.*, 2013, **3**, 3238.
18 J. Y. Li, H. Y. Lv, W. Lu, D. F. Shao, R. C. Xiao and Y. P. Sun, Phys. Lett. A, 2016,380(46),3928-3931.
19 A. Shahbazi Kootenaei and G. Ansari, *Phys. Lett. A*, 2016, **380**, 2664-2668.
20 H. Sun, Q. Li and X. G. Wan, *Phys. Chem. Chem. Phys.*, 2016, **18**, 14927-14932.
21 B. Feng, J. Zhang, Q. Zhong, W. Li, S. Li, H. Li, P. Cheng, S. Meng, L. Chen and K. Wu, *Nat. Chem.*, 2016, **8**, 563-568.
22 H. R. Jiang, Z. Lu, M. C. Wu, F. Ciucci and T. S. Zhao, *Nano Energy*, 2016, **23**, 97-104.
23 L. Shi, T. Zhao, A. Xu and J. Xu, *Sci. Bull.*, 2016, **61**, 1138-1144.
24 B. Mortazavi, A. Dianat, O. Rahaman, G. Cuniberti and T. Rabczuk, *J. Power Sources*, 2016, **329**, 456-461.
25 H. Chang, J. Cheng, X. Liu, J. Gao, M. Li, J. Li, X. Tao, F. Ding and Z. Zheng, *Chem. Eur. J.*, 2011, **17**, 8896-8903.
26 M. Mirnezhad, R. Ansari, H. Rouhi, M. Seifi and M. Faghihnasiri, *Solid State Commun.*, 2012, **152**, 1885-1889.
27 H. He and B. Pan, Eur. Phys. J. B, 2014, 87,1-6.
28 R. Ansari, M. Mirnezhad and H. Rouhi, *Nano*, 2014, **09**, 1450043.
29 B. S. Pujari, S. Gusarov, M. Brett and A. Kovalenko, *Phys. Rev. B*, 2011, **84**, 041402(R).
30 J. O. Sofo, A. S. Chaudhari and G. D. Barber, *Phys. Rev. B*, 2007, **75**, 153401.
31 S. Putz, M. Gmitra and J. Fabian, *Phys. Rev. B*, 2014, **89**, 035437.
32 D. W. Boukhvalov, M. I. Katsnelson and A. I. Lichtenstein, *Phys. Rev. B*, 2008, **77**, 035427.
33 J. Zhou, Q. Wang, Q. Sun and P. Jena, *Phys. Rev. B*, 2010, **81**, 085442.
34 A. Bhattacharya, S. Bhattacharya and G. P. Das, *Phys. Rev. B*, 2011, **84**, 075454.
35 R. Ansari, M. Mirnezhad and H. Rouhi, *Solid State Commun.*, 2015, **201**, 1-4.
36 E. Cadelano, P. L. Palla, S. Giordano and L. Colombo, *Phys. Rev. B*, 2010, **82**, 235414
37 L. Xu, A. Du and L. Kou, *ArXiv e-prints 1602*, 2016, http://arxiv.org/abs/1602.03620.
38 M. Mirnezhad, R. Ansari and H. Rouhi, *Superlattices and Microst.*, 2013, **53**, 223-231.
39 T. Y. Lü, J. C. Zheng and Y. Zhang, *Chem. Phys. Chem.*, 2015, **16**, 3015-3020.
40 T. Y. Lü, X. X. Liao, H. Q. Wang and J. C. Zheng, *J. Mater. Chem.*, 2012, **22**, 10062.
41 N. Wei, Y. Chen, K. Cai, J. H. Zhao, H. Q. Wang and J. C. Zheng, *Carbon*, 2016, **104**, 203-213.
42 M. S. Wu, B. Xu and C. Y. Ouyang, *J. Mater. Sci.*, 2016, **51**, 4691-4696.
43 H. J. Yan, Z. Q. Wang, B. Xu and C. Y. Ouyang, *Funct. Mater. Lett.*, 2012, **05**, 1250037.
44 F. H. Ning, S. Li, B. Xu and C. Y. Ouyang, *Solid State Ionics*, 2014, **263**, 46-48.
45 M. Topsakal, S. Cahangirov and S. Ciraci, *Appl. Phys. Lett.*, 2010, **96**, 091912.
46 H. J. Zhao, *Phys. Lett. A*, 2012, **376**, 3546-3550.
47 P. Giannozzi, S. Baroni, N. Bonini, M. Calandra, R. Car, C. Cavazzoni, D. Ceresoli, G. L. Chiarotti, M. Cococcioni, I. Dabo, A. Dal Corso, S. de Gironcoli, S. Fabris, G. Fratesi, R. Gebauer, U. Gerstmann, C. Gougoussis, A. Kokalj, M. Lazzeri, L. Martin-Samos, N. Marzari, F. Mauri, R. Mazzarello, S. Paolini, A. Pasquarello, L. Paulatto, C. Sbraccia, S. Scandolo, G. Sclauzero, A. P. Seitsonen, A. Smogunov, P. Umari and R. M. Wentzcovitch, *J. Phys. Condens. Matter.*, 2009, **21**, 395502.
48 D. Vanderbilt, *Phys. Rev. B*, 1990, **41**, 7892.
49 J. P. Perdew, K. Burke and M. Ernzerhof, *Phys. Rev. Lett.*, 1996, **77**, 3865.
50 R. C. Andrew, R. E. Mapasha, A. M. Ukpong and N. Chetty, *Phys. Rev. B*, 2012, **85**, 125428.
51 C. Yang, Z. Yu, P. Lu, Y. Liu, H. Ye and T. Gao, *Comput. Mater. Sci.*, 2014, **95**, 420-428.
52 H. R. Shea, R. Ramesham, C. H. Yang, Z. Y. Yu, P. F. Lu, Y. M. Liu, S. Manzoor, M. Li and S. Zhou, *Proc. SPIE 8975 id. 89750K DOI:10.1117/12.2038401*, 2014, **8975**, 89750K.
53 F. Liu, P. Ming and J. Li, *Phys. Rev. B*, 2007, **76**, 064120.
54 Q. Peng, Z. Chen and S. De, *Mech. Adv. Mater. Struct.*, 2015, **22**, 717-721.
55 B. Mohan, A. Kumar and P. K. Ahluwalia, *Physica E*, 2014, **61**, 40-47.
56 Q. Peng and S. De, *Nanoscale*, 2014, **6**, 12071-12079.
57 Q. Peng, W. Ji and S. De, *Comput. Mater. Sci.*, 2012, **56**, 11-17.
58 Q. Wei and X. Peng, *Appl. Phys. Lett.*, 2014, **104**, 251915.
59 T. Li, *Phys. Rev. B*, 2012, **85**, 235407.
60 B. Mortazavi, O. Rahaman, A. Dianat and T. Rabczuk, *Phys. Chem. Chem. Phys.*, 2016, **18**, 27405-27413.
61 B. Peng, H. Zhang, H. Z. Shao, Z. Y. Ning, Y. F. Xu, H. L. Lu, D. W. Zhang and H. Y. Zhu, *ArXiv e-prints 1602*, 2016, **1608**, https://arxiv.org/abs/1608.05877.
62 X. F. Zhou, X. Dong, A. R. Oganov, Q. Zhu, Y. J. Tian and H. T. Wang, *Phys. Rev. Lett.*, 2014, **112**, 085502.
63 J. Gao, J. Zhang, H. Liu, Q. Zhang and J. Zhao, *Nanoscale*, 2013, **5**, 9785-9792.
64 X. D. Wei, B. Fragneaud, C. A. Marianetti and J. W. Kysar, *Phys. Rev. B*, 2009, **80**, 205407